\documentclass[conference]{IEEEtran}
\IEEEoverridecommandlockouts
\usepackage{cite}
\usepackage{amsmath,amssymb,amsfonts}
\usepackage{algorithmic}
\usepackage{graphicx}
\usepackage{textcomp}
\usepackage{xcolor}
\usepackage{adjustbox}
\usepackage{multirow}
\usepackage{subcaption}
\usepackage{soul}
\usepackage[ruled,vlined]{algorithm2e}
\def\BibTeX{{\rm B\kern-.05em{\sc i\kern-.025em b}\kern-.08em
    T\kern-.1667em\lower.7ex\hbox{E}\kern-.125emX}}
\begin{document}

\author{\IEEEauthorblockN{Quincy Card, Daniel Simpson, Kshitiz Aryal, Maanak Gupta}
\IEEEauthorblockA{\textit{Department of Computer Science} \\
\textit{Tennessee Technological University}\\
Cookeville, TN USA\\
qacard42@tntech.edu, dnsimpson42@tntech.edu, karyal42@tntech.edu, mgupta@tntech.edu}
\and
\IEEEauthorblockN{Sheikh Rabiul Islam}
\IEEEauthorblockA{\textit{Department of Computer Science} \\
\textit{Rutgers University}\\
Camden, NJ USA\\
sheikh.islam@rutgers.edu}
}

\title{%Explainable Dynamic and Online Malware Classification
%Dynamic and Online Malware Classification with Explainable Methods
Explainable Deep Learning Models for \\Dynamic and Online Malware Classification
}

%\fi
\maketitle
\begin{abstract}
  %In recent years, there has been a major increase in malware activity, which has led to more need for proper response and remediation strategies. There has been research into machine learning models for malware classification, and neural networks have been shown to be effective in these tasks. These models however are not transparent and inherently explainable, requiring researchers to explore methods that can show how these models are coming to their decisions. Our paper seeks to explore explainable malware classification in the dynamic and online fields. In this paper, we use SHAP, LIME, and Permutation Importance to explain deep learning models and attribute significance to features for malware categories.
In recent years, there has been a significant surge in malware attacks, necessitating more advanced preventive measures and remedial strategies. While several successful AI-based malware classification approaches exist\textemdash{categorized into static, dynamic, or online analysis}\textemdash most successful AI models lack easily interpretable decisions and explanations for their processes. Our paper aims to delve into explainable malware classification across various execution environments (such as dynamic and online), thoroughly analyzing their respective strengths, weaknesses, and commonalities. To evaluate our approach, we train Feed Forward Neural Networks (FFNN) and Convolutional Neural Networks (CNN) to classify malware based on features obtained from dynamic and online analysis environments. The feature attribution for malware classification is performed by explainability tools, SHAP, LIME and Permutation Importance. We perform a detailed evaluation of the calculated global and local explanations from the experiments, discuss limitations and, ultimately, offer recommendations for achieving a balanced approach.
  
\end{abstract}


\section{Introduction}
Malware poses a significant cybersecurity threat, requiring effective classification for remediation~\cite{aryal2022analysis,aryal2024intra}. Machine learning-based approaches, such as those proposed by Tobiyama et al. \cite{tobiyama2016}, aim to categorize malware types for tailored response plans. Malware analysis is typically categorized into static, dynamic, and online approaches \cite{rahalilashkari2020, keyes2021, manthena2023, brown2024automated, aryal2022survey}. Static analysis examines resting malicious files, while dynamic analysis executes malware in a simulated environment. Online analysis, on the other hand, monitors systems in real-time, potentially accessing internet resources. Both physical test beds \cite{karapoola2022} and cloud environments \cite{manthena2023} are used for this purpose, although costs can be prohibitive.

Despite the importance of accurate malware classification, understanding the classification processes is equally crucial. Deep learning models, often used for their accuracy, can be challenging to interpret due to their complexity \cite{hall2019}. Explainable AI/ML methods aim to address this challenge by elucidating decisions and identifying significant features contributing to model outcomes. This enhances user trust and aids security analysts in countering malware threats \cite{blancojusticia2019}.

In this work, we apply explainable techniques for feature attribution of deep learning models used in dynamic and online malware classification environments. The first phase of work deals with training the deep learning models to create a malware detector on dynamic and online analysis features. A Feed Forward Neural Network (FFNN) and Convolutional Neural Network (CNN) were trained for both data sets. The second phase of the research involves employing Shapley Additive exPlanations (SHAP), Local Interpretable Model-Agnostic Explanations (LIME), and Permutation Importance to interpret black-box model decisions. SHAP \cite{lundberg2017} assigns contributions of each feature to predictions based on cooperative game theory. LIME \cite{ribeiro2016} constructs local interpretable models around specific predictions by generating perturbed instances. Permutation Importance \cite{breiman2001} assesses feature impact by shuffling values, offering global explanations. We use DeepSHAP \cite{lundberg2017} as it provides both global and local interpretations, complemented by Permutation Importance for global explanations and LIME for local insights. The main contributions of this paper include:
\begin{itemize}
\item We evaluate the effectiveness of deep learning models for classifying malware categories from both a dynamic and an online data set.
\item We extend this analysis by explaining model predictions on a global level with SHAP and Permutation Importance, and on a local level with LIME and SHAP.
\end{itemize}

The paper is organized as follows. Section \ref{sec:background} reviews related works in (a) dynamic analysis, (b) online analysis, and (c) explainability techniques for interpreting model predictions. Section \ref{sec:methodology} outlines the methodology and introduces the dynamic and online datasets. Results and model explanations for each dataset are presented in Section \ref{sec:results}. The paper concludes with a summary and discussion of future work in Section \ref{sec:conclusion}.

\begin{table*}
\caption{Related works compared. A $\surd$ indicates that a specific paper has this feature or model, and a blank cell shows that this attribute or model does not exist.}
\label{tab:related-works}
\begin{adjustbox}{width=\textwidth}
\begin{tabular}{|l|lcl|cl|lllll|llclllc|lllllll|cll|}
\hline
                            & \multicolumn{3}{c|}{\textbf{Analysis}}                                                                                & \multicolumn{2}{c|}{\textbf{Domain}}                                          & \multicolumn{5}{c|}{\textbf{Features}}                                                                                                                                                        & \multicolumn{7}{c|}{\textbf{Models Used}}                                                                                                                                                                                                                                     & \multicolumn{7}{c|}{\textbf{XAI Technique}}                                                                                                                                                                                                   & \multicolumn{3}{c|}{\textbf{Platform}}                                                              \\ \hline
\multicolumn{1}{|c|}{\multirow{-9}{*}{\textbf{Paper}}} & \rotatebox{90}{Detection} & \multicolumn{1}{|c|}{\rotatebox{90}{Category Classification}} & \rotatebox{90}{Family Classification}        & \multicolumn{1}{|c|}{\rotatebox{90}{Dynamic Analysis}} & \rotatebox{90}{Online Analysis}              & \rotatebox{90}{Performance Metrics} & \multicolumn{1}{|c|}{\rotatebox{90}{API Calls}} & \rotatebox{90}{System Calls} & \multicolumn{1}{|c|}{\rotatebox{90}{Grayscale Images}} & \rotatebox{90}{Others}                       & \rotatebox{90}{Logistic Regression} & \multicolumn{1}{|c|}{\rotatebox{90}{Decision Tree}} & \rotatebox{90}{Random Forest} & \multicolumn{1}{|c|}{\rotatebox{90}{LightGBM}} & \rotatebox{90}{XGBoost} & \multicolumn{1}{|c|}{\rotatebox{90}{Support Vector Machine}} & \rotatebox{90}{Deep Learning} & \rotatebox{90}{SHAP}    & \multicolumn{1}{|c|}{\rotatebox{90}{LIME}}    & \rotatebox{90}{LRP}     & \multicolumn{1}{|c|}{\rotatebox{90}{Grad-CAM}} & \rotatebox{90}{Heatmap} & \multicolumn{1}{|c|}{\rotatebox{90}{Integrated Gradients}} & \rotatebox{90}{Permutation Importance}       & \rotatebox{90}{Linux}   & \multicolumn{1}{|c|}{\rotatebox{90}{Android}} & \rotatebox{90}{Windows}                      \\ \hline
Schofield et al. (2021) \cite{schofield2021}           & \multicolumn{1}{l|}{}          & \multicolumn{1}{c|}{$\surd$}                 &                              & \multicolumn{1}{c|}{$\surd$}          &                              & \multicolumn{1}{l|}{}                    & \multicolumn{1}{c|}{$\surd$}   & \multicolumn{1}{l|}{}             & \multicolumn{1}{l|}{}                 &                              & \multicolumn{1}{l|}{}                    & \multicolumn{1}{c|}{$\surd$}       & \multicolumn{1}{c|}{$\surd$}       & \multicolumn{1}{l|}{}         & \multicolumn{1}{l|}{}        & \multicolumn{1}{l|}{}                       & $\surd$                            & \multicolumn{1}{l|}{}        & \multicolumn{1}{l|}{}        & \multicolumn{1}{l|}{}        & \multicolumn{1}{l|}{}         & \multicolumn{1}{l|}{}        & \multicolumn{1}{l|}{}                     &                              & \multicolumn{1}{c|}{}        & \multicolumn{1}{l|}{}        & \multicolumn{1}{c|}{$\surd$} \\ \hline
Keyes et al. (2021) \cite{keyes2021}              & \multicolumn{1}{l|}{}          & \multicolumn{1}{c|}{$\surd$}                 &                              & \multicolumn{1}{c|}{$\surd$}          &                              & \multicolumn{1}{l|}{}                    & \multicolumn{1}{c|}{$\surd$}   & \multicolumn{1}{l|}{}             & \multicolumn{1}{l|}{}                 & \multicolumn{1}{c|}{$\surd$} & \multicolumn{1}{l|}{}                    & \multicolumn{1}{c|}{$\surd$}       & \multicolumn{1}{c|}{$\surd$}       & \multicolumn{1}{l|}{}         & \multicolumn{1}{l|}{}        & \multicolumn{1}{l|}{}                       & \multicolumn{1}{l|}{}              & \multicolumn{1}{l|}{}        & \multicolumn{1}{l|}{}        & \multicolumn{1}{l|}{}        & \multicolumn{1}{l|}{}         & \multicolumn{1}{l|}{}        & \multicolumn{1}{l|}{}                     &                              & \multicolumn{1}{c|}{}        & \multicolumn{1}{c|}{$\surd$} &                              \\ \hline
Pirch et al. (2021) \cite{pirch2021}              & \multicolumn{1}{l|}{}          & \multicolumn{1}{l|}{}                        & \multicolumn{1}{c|}{$\surd$} & \multicolumn{1}{c|}{$\surd$}          &                              & \multicolumn{1}{l|}{}                    & \multicolumn{1}{l|}{}          & \multicolumn{1}{l|}{}             & \multicolumn{1}{l|}{}                 & \multicolumn{1}{c|}{$\surd$} & \multicolumn{1}{l|}{}                    & \multicolumn{1}{l|}{}              & \multicolumn{1}{l|}{}              & \multicolumn{1}{l|}{}         & \multicolumn{1}{l|}{}        & \multicolumn{1}{l|}{}                       & $\surd$                            & \multicolumn{1}{l|}{}        & \multicolumn{1}{l|}{}        & \multicolumn{1}{c|}{$\surd$} & \multicolumn{1}{l|}{}         & \multicolumn{1}{l|}{}        & \multicolumn{1}{l|}{}                     &                              & \multicolumn{1}{c|}{}        & \multicolumn{1}{l|}{}        &         $\surd$                     \\ \hline
Alenezi et al. (2021) \cite{alenezi2021}        & \multicolumn{1}{l|}{}          & \multicolumn{1}{c|}{$\surd$}                 &                              & \multicolumn{1}{c|}{$\surd$}          &                              & \multicolumn{1}{l|}{}                    & \multicolumn{1}{l|}{}          & \multicolumn{1}{c|}{$\surd$}      & \multicolumn{1}{l|}{}                 &                              & \multicolumn{1}{l|}{}                    & \multicolumn{1}{l|}{}              & \multicolumn{1}{c|}{$\surd$}       & \multicolumn{1}{l|}{}         & \multicolumn{1}{c|}{$\surd$} & \multicolumn{1}{l|}{}                       & $\surd$                            & \multicolumn{1}{c|}{$\surd$} & \multicolumn{1}{l|}{}        & \multicolumn{1}{l|}{}        & \multicolumn{1}{l|}{}         & \multicolumn{1}{l|}{}        & \multicolumn{1}{l|}{}                     &                              & \multicolumn{1}{c|}{}        & \multicolumn{1}{c|}{$\surd$} &                              \\ \hline
Kimmel et al. (2021) \cite{kimmell2021}             & \multicolumn{1}{c|}{$\surd$}   & \multicolumn{1}{l|}{}                        &                              & \multicolumn{1}{l|}{}                 & \multicolumn{1}{c|}{$\surd$} & \multicolumn{1}{c|}{$\surd$}             & \multicolumn{1}{l|}{}          & \multicolumn{1}{l|}{}             & \multicolumn{1}{l|}{}                 &                              & \multicolumn{1}{l|}{}                    & \multicolumn{1}{l|}{}              & \multicolumn{1}{c|}{$\surd$}       & \multicolumn{1}{l|}{}         & \multicolumn{1}{l|}{}        & \multicolumn{1}{c|}{$\surd$}                & $\surd$                            & \multicolumn{1}{l|}{}        & \multicolumn{1}{l|}{}        & \multicolumn{1}{l|}{}        & \multicolumn{1}{l|}{}         & \multicolumn{1}{l|}{}        & \multicolumn{1}{l|}{}                     &                              & \multicolumn{1}{c|}{$\surd$} & \multicolumn{1}{l|}{}        &                              \\ \hline
Prasse et al. (2021) \cite{prasse2021}             & \multicolumn{1}{l|}{}          & \multicolumn{1}{c|}{$\surd$}                 & \multicolumn{1}{c|}{$\surd$} & \multicolumn{1}{l|}{}                 & \multicolumn{1}{c|}{$\surd$} & \multicolumn{1}{l|}{}                    & \multicolumn{1}{l|}{}          & \multicolumn{1}{l|}{}             & \multicolumn{1}{l|}{}                 & \multicolumn{1}{c|}{$\surd$} & \multicolumn{1}{l|}{}                    & \multicolumn{1}{l|}{}              & \multicolumn{1}{c|}{$\surd$}       & \multicolumn{1}{l|}{}         & \multicolumn{1}{l|}{}        & \multicolumn{1}{c|}{}                       & $\surd$                            & \multicolumn{1}{l|}{}        & \multicolumn{1}{l|}{}        & \multicolumn{1}{l|}{}        & \multicolumn{1}{l|}{}         & \multicolumn{1}{l|}{}        & \multicolumn{1}{c|}{$\surd$}              &                              & \multicolumn{1}{c|}{}        & \multicolumn{1}{l|}{}        & \multicolumn{1}{c|}{$\surd$} \\ \hline
Karn et al. (2021) \cite{karn2021}               & \multicolumn{1}{c|}{$\surd$}   & \multicolumn{1}{l|}{}                        &                              & \multicolumn{1}{l|}{}                 & \multicolumn{1}{c|}{$\surd$} & \multicolumn{1}{l|}{}                    & \multicolumn{1}{l|}{}          & \multicolumn{1}{c|}{$\surd$}      & \multicolumn{1}{l|}{}                 &                              & \multicolumn{1}{l|}{}                    & \multicolumn{1}{c|}{$\surd$}       & \multicolumn{1}{l|}{}              & \multicolumn{1}{l|}{}         & \multicolumn{1}{c|}{$\surd$} & \multicolumn{1}{l|}{}                       & $\surd$                            & \multicolumn{1}{c|}{$\surd$} & \multicolumn{1}{c|}{$\surd$} & \multicolumn{1}{l|}{}        & \multicolumn{1}{l|}{}         & \multicolumn{1}{l|}{}        & \multicolumn{1}{l|}{}                     &                              & \multicolumn{1}{c|}{$\surd$}        & \multicolumn{1}{l|}{}        &                              \\ \hline
Ullah et al. (2022) \cite{ullah2022}              & \multicolumn{1}{l|}{}          & \multicolumn{1}{c|}{$\surd$}                 &                              & \multicolumn{1}{c|}{$\surd$}          &                              & \multicolumn{1}{l|}{}                    & \multicolumn{1}{l|}{}          & \multicolumn{1}{l|}{}             & \multicolumn{1}{c|}{$\surd$}          & \multicolumn{1}{c|}{$\surd$} & \multicolumn{1}{c|}{$\surd$}             & \multicolumn{1}{c|}{$\surd$}       & \multicolumn{1}{c|}{$\surd$}       & \multicolumn{1}{l|}{}         & \multicolumn{1}{l|}{}        & \multicolumn{1}{c|}{$\surd$}                & \multicolumn{1}{l|}{}              & \multicolumn{1}{c|}{$\surd$} & \multicolumn{1}{l|}{}        & \multicolumn{1}{l|}{}        & \multicolumn{1}{l|}{}         & \multicolumn{1}{l|}{}        & \multicolumn{1}{l|}{}                     &                              & \multicolumn{1}{c|}{}        & \multicolumn{1}{c|}{$\surd$} &                              \\ \hline
Naeem et al. (2022) \cite{naeem2022}              & \multicolumn{1}{l|}{}          & \multicolumn{1}{l|}{}                        & \multicolumn{1}{c|}{$\surd$} & \multicolumn{1}{c|}{$\surd$}          &                              & \multicolumn{1}{l|}{}                    & \multicolumn{1}{l|}{}          & \multicolumn{1}{l|}{}             & \multicolumn{1}{c|}{$\surd$}          &                              & \multicolumn{1}{c|}{$\surd$}             & \multicolumn{1}{c|}{$\surd$}       & \multicolumn{1}{c|}{$\surd$}       & \multicolumn{1}{l|}{}         & \multicolumn{1}{l|}{}        & \multicolumn{1}{c|}{$\surd$}                & $\surd$                            & \multicolumn{1}{l|}{}        & \multicolumn{1}{l|}{}        & \multicolumn{1}{l|}{}        & \multicolumn{1}{c|}{$\surd$}  & \multicolumn{1}{c|}{$\surd$} & \multicolumn{1}{l|}{}                     &                              & \multicolumn{1}{c|}{}        & \multicolumn{1}{c|}{$\surd$} &                              \\ \hline
Brown et al. (2022) \cite{brown2022}               & \multicolumn{1}{l|}{}          & \multicolumn{1}{c|}{$\surd$}                 &                              & \multicolumn{1}{l|}{}                 & \multicolumn{1}{c|}{$\surd$} & \multicolumn{1}{l|}{}                    & \multicolumn{1}{l|}{}          & \multicolumn{1}{c|}{$\surd$}      & \multicolumn{1}{l|}{}                 &                              & \multicolumn{1}{l|}{}                    & \multicolumn{1}{l|}{}              & \multicolumn{1}{l|}{}              & \multicolumn{1}{c|}{$\surd$}  & \multicolumn{1}{l|}{}        & \multicolumn{1}{l|}{}                       & \multicolumn{1}{l|}{}              & \multicolumn{1}{l|}{}        & \multicolumn{1}{l|}{}        & \multicolumn{1}{l|}{}        & \multicolumn{1}{l|}{}         & \multicolumn{1}{l|}{}        & \multicolumn{1}{l|}{}                     &                              & \multicolumn{1}{c|}{$\surd$} & \multicolumn{1}{l|}{}        &                              \\ \hline
Manthena et al. (2023) \cite{manthena2023}            & \multicolumn{1}{c|}{$\surd$}   & \multicolumn{1}{l|}{}                        &                              & \multicolumn{1}{l|}{}                 & \multicolumn{1}{c|}{$\surd$} & \multicolumn{1}{c|}{$\surd$}             & \multicolumn{1}{l|}{}          & \multicolumn{1}{l|}{}             & \multicolumn{1}{l|}{}                 &                              & \multicolumn{1}{l|}{}                    & \multicolumn{1}{l|}{}              & \multicolumn{1}{c|}{$\surd$}       & \multicolumn{1}{l|}{}         & \multicolumn{1}{l|}{}        & \multicolumn{1}{c|}{$\surd$}                & $\surd$                            & \multicolumn{1}{c|}{$\surd$} & \multicolumn{1}{l|}{}        & \multicolumn{1}{l|}{}        & \multicolumn{1}{l|}{}         & \multicolumn{1}{l|}{}        & \multicolumn{1}{l|}{}                     &                              & \multicolumn{1}{c|}{$\surd$}        & \multicolumn{1}{l|}{}        &                              \\ \hline
{\bfseries Our Approach}                & \multicolumn{1}{l|}{}          & \multicolumn{1}{c|}{$\surd$}                 &                              & \multicolumn{1}{c|}{$\surd$}          & \multicolumn{1}{c|}{$\surd$} & \multicolumn{1}{l|}{}                    & \multicolumn{1}{c|}{$\surd$}   & \multicolumn{1}{l|}{}             & \multicolumn{1}{l|}{}                 & \multicolumn{1}{c|}{$\surd$} & \multicolumn{1}{l|}{}                    & \multicolumn{1}{c|}{}              & \multicolumn{1}{l|}{}              & \multicolumn{1}{l|}{}         & \multicolumn{1}{l|}{}        & \multicolumn{1}{l|}{}                       & $\surd$                            & \multicolumn{1}{c|}{$\surd$} & \multicolumn{1}{c|}{$\surd$} & \multicolumn{1}{l|}{}        & \multicolumn{1}{l|}{}         & \multicolumn{1}{l|}{}        & \multicolumn{1}{l|}{}                     & \multicolumn{1}{c|}{$\surd$} & \multicolumn{1}{c|}{}        & \multicolumn{1}{c|}{$\surd$} &  \multicolumn{1}{c|}{$\surd$}                             \\ \hline
\end{tabular}
\end{adjustbox}
\end{table*}

\section{Related Works}\label{sec:background}
This section provides a review of dynamic and online analyses, concluding with a discussion on explainable AI. Table \ref{tab:related-works} compares relevant works across attributes like domain, analysis level, models, features, explainability techniques, and malware platform. It's important to clarify terms often conflated in the literature. Detection involves determining malware's presence or absence, similar to binary classification. This work distinguishes detection from classification, where the goal is to differentiate malware samples. Classification groups malware by family or category, with family denoting variants sharing traits and category grouping malware by objectives (e.g., ransomware encrypting systems or files).

\subsection{Dynamic Analysis}
Several studies focus on dynamic malware analysis in traditional host-based environments. \cite{keyes2021} conducted Android malware classification using diverse input features, including memory, API calls, network data, battery status, log writing and process count. Transparent models like Naive Bayes and Decision Tree were employed, as a well as black-box SVM model. \cite{schofield2021} classified Windows malware categories using API calls and various transparent and deep learning models, such as a CNN. Despite advancements over static analysis, these approaches also face challenges in analyzing modern advanced malware that may detect them in a closed environment and evade analysis, prompting the need for online malware analysis methods.

\subsection{Online Analysis}
Several studies focus on online malware analysis, addressing the limitations of dynamic analysis by continuously monitoring systems and utilizing various runtime features for machine learning models. In \cite{kimmell2021}, performance metrics were employed to detect malware with six ML classifiers, including a CNN. \cite{brown2022} categorized Linux malware in a cloud environment using the LightGBM model trained on system call n-grams. These studies share the characteristic of conducting online analysis in a cloud environment, chosen for its relative resource efficiency as compared to traditional host-based environments. Almost all of the previously mentioned works used some black-box model but lacked the necessary post-hoc explanations. Notably, our work stands out by combining dynamic and online analysis in a non-cloud, host-based environment, along with explainability methods for interpreting deep learning models in malware classification.

\subsection{Explainable AI}
Some machine learning models, termed "transparent models," operate without the need for external or post-hoc explanation methods. Typically, these models, such as tree-based models, offer easy visualization through their tree structure. Conversely, approaches like SVM or neural networks are labeled as black box methods, where the internal mechanisms are not readily visible or understandable, necessitating additional post-hoc explanations. These explanations can be categorized based on their locality and model-specificity. Local explanations focus on understanding the model's predictions within specific examples, such as classifying individual pixels in an image. In contrast, global explanations provide insights into general patterns, feature importance, and model structure without examining specific inputs \cite{lin2021interpreting}. There may be scenarios where only a global or local explanation suffices, but generally, both levels of locality are essential for model interpretability. Model-specific techniques are tailored to a particular model type, such as saliency maps for CNNs, while model-agnostic techniques are applicable across different architectures.

Several studies address machine learning model interpretability in malware analysis. In \cite{pirch2021}, an interpretable CNN model predicted tags for dynamic malware family categorization, with Layer-wise Relevance Propagation (LRP) used for explanation. \cite{naeem2022} developed a dynamic Android malware classification method, employing models such as Random Forest, Decision Tree, SVM, CNN, and Logistic Regression on grayscale images, explained through local, model-specific methods like Grad-CAM and heatmaps. \cite{alenezi2021} and \cite{ullah2022} explored SHAP for explaining malware classification, focusing on dynamic Android malware classification and transformers-based transfer learning, respectively. \cite{prasse2021} classified online Windows malware using models like Random Forest, LSTM, CNN, and Transformer, explained by Integrated Gradients for global and local insights. \cite{manthena2023} focued on real-time cloud-based malware detection, explaining predictions with SHAP for various black-box models. Karn et al. \cite{karn2021} explain online malware detection, particularly cryptominer detection using SHAP for XGBoost predictions and LIME for other models. While these studies emphasize online analysis and explainability, none specifically address explaining both dynamic and online malware category classification in a traditional host-based environment.

\section{Methodology}\label{sec:methodology}
In this section, we discuss the methodology we use for our dynamic analysis, our online analysis, how we evaluate our models, and our approach to explainability.

\subsection{Dynamic Analysis}
We utilized the AndMal2020 dataset from the Canadian Center for Cybersecurity \cite{cccsc2020}, comprising 12 Android malware categories, each with 141 features across 6 types, including memory, API calls, network, battery, log writing, and total process count. Table \ref{tab:dynamic-distribution} shows the class distribution is highly imbalanced, and we found excluding minority classes and adjusting class weights to be ineffective. SMOTE (Synthetic Minority Oversampling Technique) was most effective, balancing the dataset by generating synthetic samples between the feature space of real examples. For both analyses, we employed a FFNN and a CNN, providing a simpler and a more complex deep learning model, respectively. The FFNN consisted of 6 hidden layers, including 2 fully connected layers and 1 dropout layer, all using \textit{ReLU} activation, with \textit{Softmax} for the output layer. Trained on an 80\% training dataset for 135 epochs with a batch size of 10, testing was conducted on the remaining 20\%. The CNN architecture included a convolution layer, max pooling layer, and two fully connected layers with dropout. Training occurred on 80\% of the data for 75 epochs with a batch size of 10, with testing on the remaining 20\%.

\begin{table}
    \caption{All classes of Dynamic data set}
    \label{tab:dynamic-distribution}
    \centering
    \begin{tabular}{|c|c|}
    \hline
        \textbf{Category} & \textbf{Number of Samples} \\
        \hline
        Riskware & 7261         \\ \hline
        Adware & 5838           \\ \hline
        Trojan & 4412           \\ \hline
        Ransomware & 1861       \\ \hline
        Trojan\_Spy & 1801      \\ \hline
        Trojan\_SMS & 1028      \\ \hline
        Trojan\_Dropper & 837   \\ \hline
        PUA & 665               \\ \hline
        Backdoor & 591          \\ \hline
        Scareware & 462         \\ \hline
        FileInfector & 129      \\ \hline
        Trojan\_Banker & 118    \\
        \hline
    \end{tabular}
\end{table}

\subsection{Online Analysis}
We used the RaDaR dataset from the Indian Institute of Technology Madras \cite{karapoola2022}, capturing real-time behavior of Windows malware on a physical testbed. This dataset facilitates analysis of modern malware capable of detecting sandbox environments and remaining dormant. However, the extensive resources required for this analysis, along with the larger data volume generated, result in increased computation times for classification and explainability methods. The dataset comprises five malware categories, with 55 initial features focusing on malware behavior at the hardware level. Like dynamic analysis, the online analysis dataset is highly imbalanced, as seen in Table \ref{tab:online-distribution}, which is addressed using SMOTE to create synthetic samples. Initially considering a Long Short-Term Memory (LSTM) model for its advantages in handling time-series data, we found SHAP less compatible with LSTM's data shape. Hence, we opted for models consistent across both analysis levels. Future work should explore models adept at handling time-series data and suitable explanation methods. For the CNN, the most effective hyperparameters included a convolution layer followed by max pooling, another convolution layer, max pooling, and two fully connected layers. \textit{ReLU} activation was used for all hidden layers, and \textit{Softmax} for the output layer. The model was trained on 80\% of the dataset for 75 epochs with a batch size of 50. Similarly, the FFNN comprised 5 hidden layers, with \textit{ReLU} activation, 2 fully connected layers, 1 dropout layer, and \textit{Softmax} activation for the output layer. Trained on 80\% of the dataset for 100 epochs with a batch size of 50, both models were tested on the remaining 20\% of the online dataset.

\begin{table}
    \caption{All classes of Online data set}
    \label{tab:online-distribution}
    \centering
    \begin{tabular}{|c|c|}
    \hline
        \textbf{Category} & \textbf{Number of Snapshots}  \\
        \hline
        Cryptominer & 158158            \\ \hline
        Deceptor & 99099                \\ \hline
        Ransomware & 13013              \\ \hline
        PUA & 3003                      \\ \hline
        Backdoor & 1001                 \\
        \hline
    \end{tabular}
\end{table}

\begin{figure*}
  \centering

  \subfloat{\includegraphics[width=0.5\textwidth]{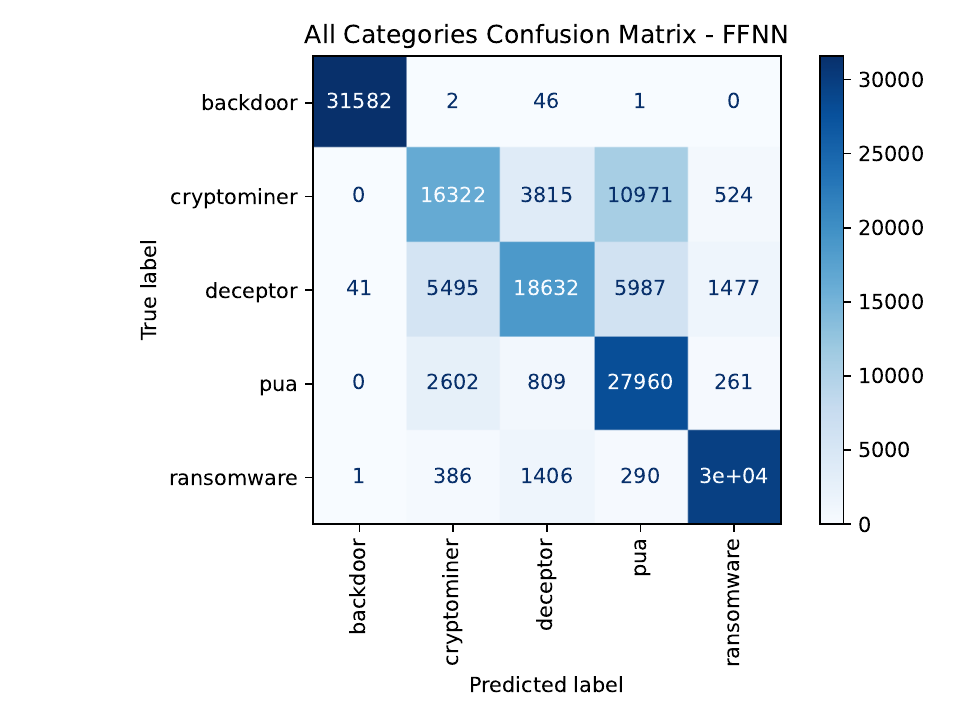}}\hfill
  \subfloat{\includegraphics[width=0.5\textwidth]{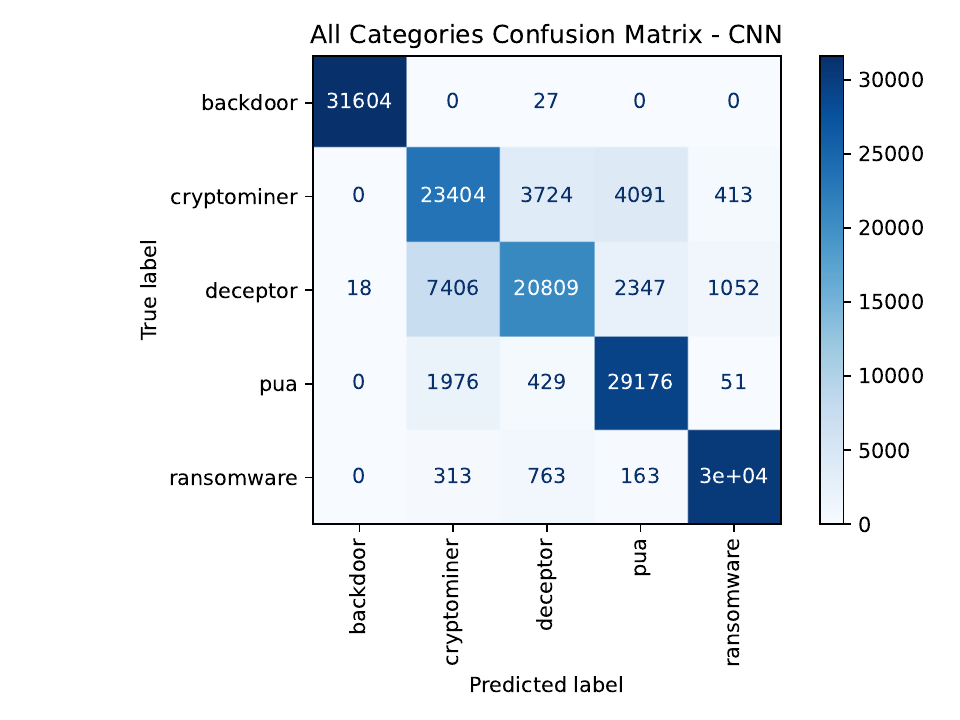}}

  \caption{Performance of models in Online Analysis}
  \label{fig:online_all_categories}
\end{figure*}

\begin{figure*}
  \centering

  \subfloat{\includegraphics[width=0.5\textwidth]{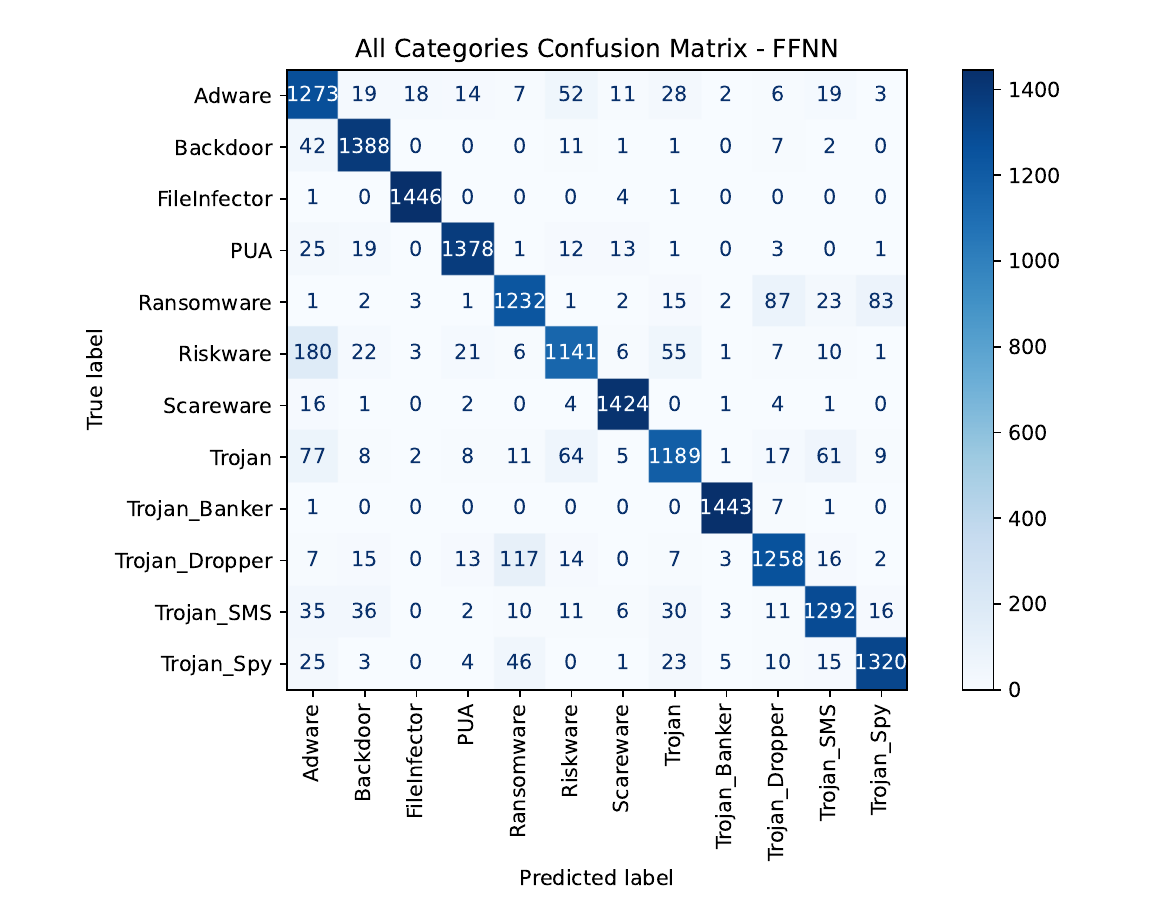}}\hfill
  \subfloat{\includegraphics[width=0.5\textwidth]{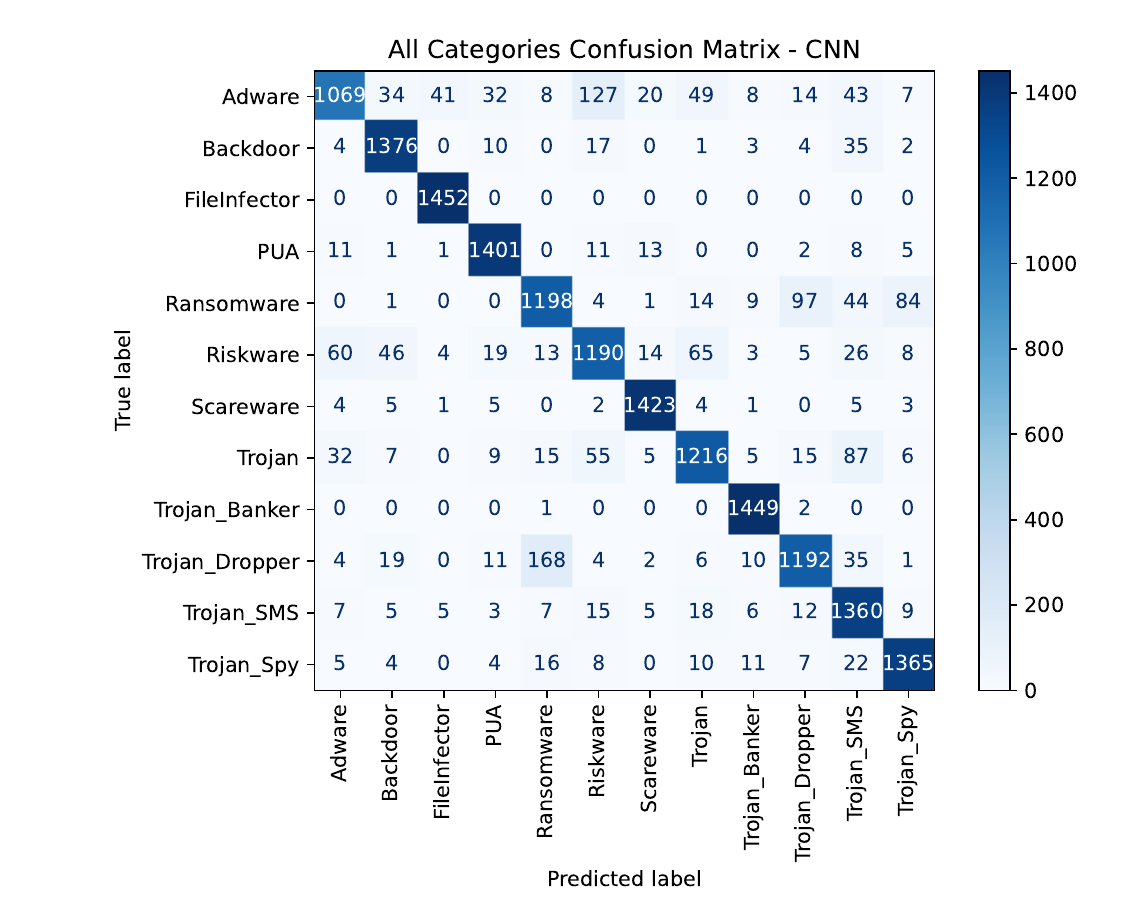}}

  \caption{Performance of models in Dynamic Analysis}
  \label{fig:dynamic_all_categories_cm}
\end{figure*}

\subsection{Explainability Approach}
We chose SHAP as our primary explanation method due to its robust and well-documented Python library. SHAP quantifies feature contributions through Shapley values, providing a mathematical framework to explain model predictions \cite{lundberg2017}. SHAP offers model-specific explanation methods, such as \textit{DeepExplainer} for deep learning models, and provides explanations at both global and local levels. After model training, we generate an explainer using SHAP's \textit{DeepExplainer}, computing SHAP values for 500 samples from the test datasets of both dynamic and online analyses. This under-sampling is necessary due to the resource-intensive nature of computing SHAP values, especially for complex architectures like CNNs.

LIME and Permutation Importance serve as supplementary explanation methods. LIME offers local explanations, while Permutation Importance provides global explanations. We chose these methods to ensure robust interpretations of the models and to verify the efficiency of SHAP under random sampling. For Permutation Importance, we permuted each feature 30 times for both dynamic and online datasets. For local explanations, we randomly selected correctly classified and misclassified observations from the subsample and generated graphs using SHAP and LIME.

\begin{table}[]
\caption{Performance Metrics for Dynamic Analysis and Riskware-Specific Metrics}
\label{tab:dynamic-performance}
\begin{adjustbox}{width=0.47\textwidth}
\begin{tabular}{|c|c|c|c|c|}
\hline
                            & Accuracy (\%) & Precision (\%) & Recall (\%) & F1 (\%) \\ \hline
FFNN without SMOTE          & 80.76         & 81.23          & 80.76       & 81.00   \\ \hline
FFNN with SMOTE             & 90.57         & 91.00          & 90.57       & 90.63   \\ \hline
CNN without SMOTE           & 81.68         & 81.79          & 81.68       & 81.74   \\ \hline
CNN with SMOTE              & 90.04         & 90.03          & 90.04       & 90.03   \\ \hline
Riskware - FNN with SMOTE   & 78.53         & 100.00         & 78.53       & 87.97   \\ \hline
Riskware - CNN with SMOTE   & 81.90         & 100.00         & 81.90       & 90.04   \\ \hline
\end{tabular}
\end{adjustbox}
\end{table}

\begin{table}[]
\caption{Performance Metrics for Online Analysis and Deceptor-Specific Metrics}
\label{tab:online-performance}
\begin{adjustbox}{width=0.46\textwidth}
\begin{tabular}{|c|c|c|c|c|}
\hline
                          & Accuracy (\%) & Precision (\%) & Recall (\%) & F1 (\%) \\ \hline
FFNN without SMOTE        & 77.45         & 77.15          & 77.45       & 77.30   \\ \hline
FFNN with SMOTE           & 78.43         & 79.16          & 78.43       & 78.79   \\ \hline
CNN without SMOTE         & 79.72         & 80.50          & 79.72       & 80.11   \\ \hline
CNN with SMOTE            & 85.60         & 85.65          & 85.60       & 85.63   \\ \hline
Deceptor - FNN with SMOTE & 58.90         & 100.00         & 58.90       & 74.14   \\ \hline
Deceptor - CNN with SMOTE & 65.78         & 100.00         & 65.78       & 79.36   \\ \hline
\end{tabular}
\end{adjustbox}
\end{table}

\section{Results and Discussion}\label{sec:results}

\begin{figure*}
  \centering

  \subfloat{\includegraphics[width=0.5\textwidth]{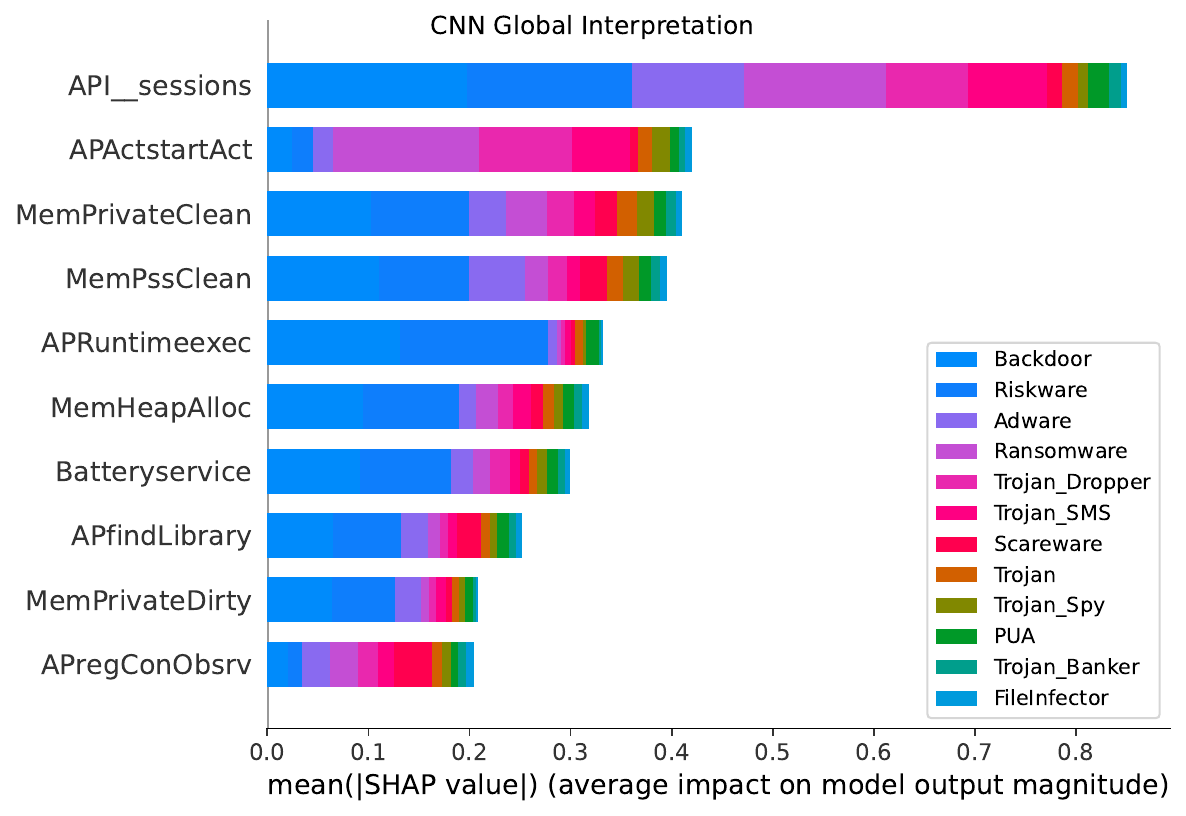}}\hfill
  \subfloat{\includegraphics[width=0.5\textwidth]{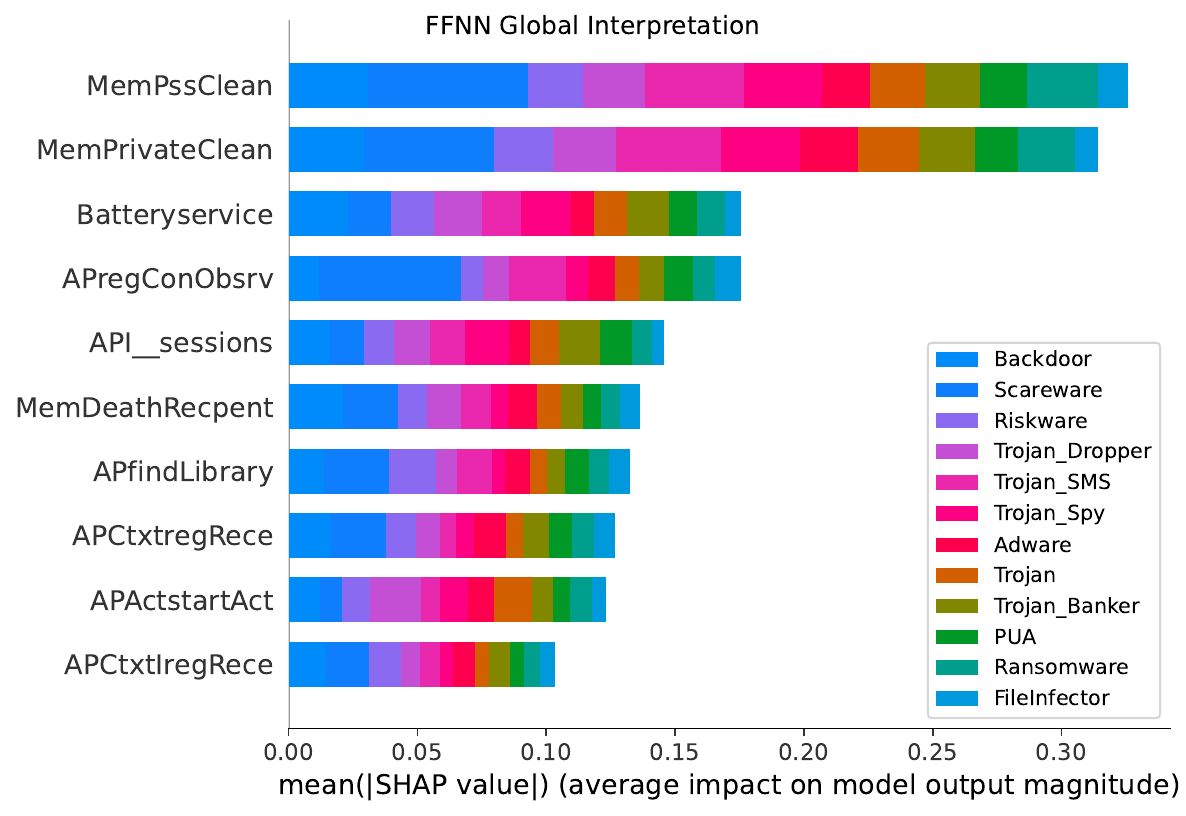}}

  \caption{A stacked bar graph depicting the top 10 online features identified by SHAP in model decision making}
  \label{fig:dynamic_global_shap}
\end{figure*}

\begin{figure*}
  \centering

  \subfloat{\includegraphics[width=0.5\textwidth]{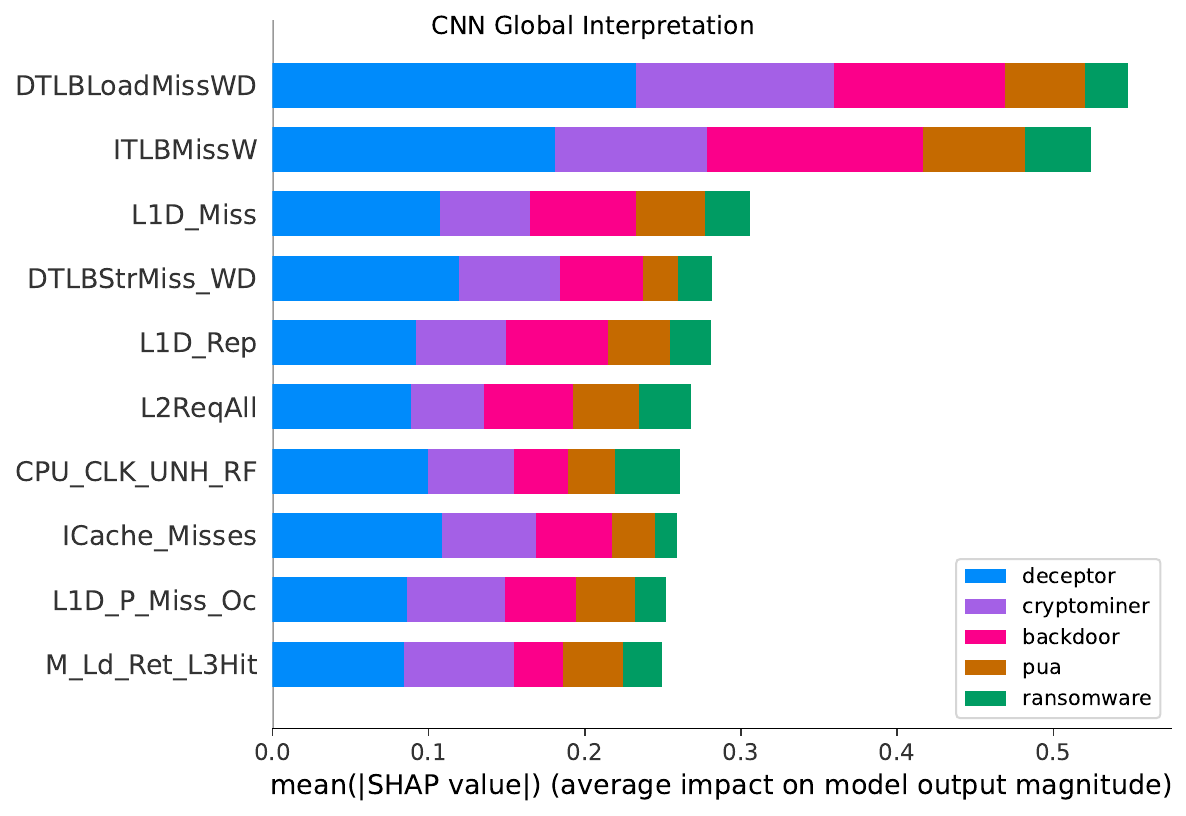}}\hfill
  \subfloat{\includegraphics[width=0.5\textwidth]{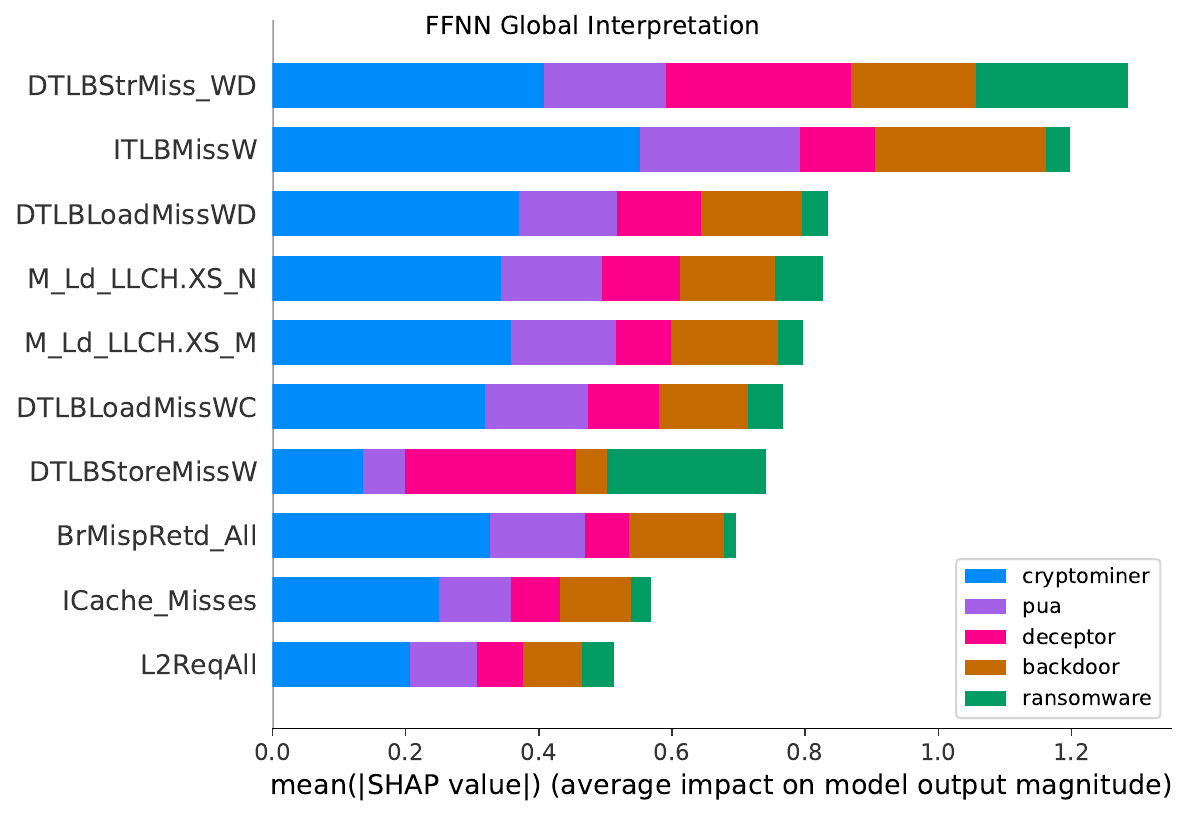}}

  \caption{A stacked bar graph depicting the top 10 online features identified by SHAP in model decision making}
  \label{fig:online_global_shap}
\end{figure*}

\begin{table}[]
\centering
\caption{Top 10 Dynamic Features Identified by Permutation Importance}
\label{tab:dynamic-permutation-importance}
\begin{tabular}{|c|c|}
\hline
\textbf{FFNN Features (Importance)} & \textbf{CNN Features (Importance)} \\ \hline
MemPssClean (0.236)        & APCtxtregRece (0.178)     \\ \hline
MemPrivateClean (0.191)    & APregConObsrv (0.158)     \\ \hline
APregConObsrv (0.140)      & MemPssClean (0.155)       \\ \hline
APCtxtregRece (0.115)      & MemPrivateClean (0.132)   \\ \hline
Batteryservice (0.110)     & APfindLibrary (0.105)     \\ \hline
MemDeathRecpent (0.108)    & MemDeathRecpent (0.103)   \\ \hline
APCtxtIregRece (0.103)     & API\_\_sessions (0.083)   \\ \hline
APActstartAct (0.102)      & APActstartAct (0.078)     \\ \hline
API\_\_sessions (0.086)    & MemParcelCount (0.072)    \\ \hline
APfindLibrary (0.079)      & MemHeapAlloc (0.069)      \\ \hline
\end{tabular}
\end{table}

\begin{table}[]
\centering
\caption{Top 10 Online Features Identified by Permutation Importance}
\label{tab:online-permutation-importance}
\begin{tabular}{|c|c|}
\hline
\textbf{FFNN Features (Importance)} & \textbf{CNN Features (Importance)} \\ \hline
L1D\_P\_Miss\_Oc (0.150)   & ICache\_Misses (0.188)    \\ \hline
DTLBLoadMissWD (0.113)     & DTLBLoadMissWD (0.181)    \\ \hline
DTLBStoreMissW (0.112)     & ITLBMissW (0.143)         \\ \hline
DTLBLoadMiss\_W (0.097)    & L2ReqAll (0.138)          \\ \hline
ITLBMissW (0.096)          & M\_Ld\_LLCH.XS\_N (0.132) \\ \hline
L2ReqPFms (0.090)          & L1D\_P\_Miss\_Oc (0.131)  \\ \hline
Core\_cyc (0.086)          & L1D\_Rep (0.130)          \\ \hline
L1D\_Rep (0.080)           & Ref\_cyc (0.103)          \\ \hline
DTLBLoadMissWC (0.077)     & L2ReqPFms (0.098)         \\ \hline
BrMispRetd\_All (0.069)    & L1D\_Rep (0.097)          \\ \hline
\end{tabular}
\end{table}

\subsection{Evaluation of Performance Metrics}
Table \ref{tab:dynamic-performance} provides performance metrics for overall model performance in dynamic analysis, including a breakdown for one of the worst-performing classes. Utilizing the F1 score, a comprehensive evaluation beyond accuracy, the FFNN model achieved 90.63\%, while the CNN model reached 90.03\%. The comparison with and without SMOTE intervention indicates a notable performance increase with synthetic samples. Despite Riskware being the majority class in the unaltered dataset, its relative poor performance suggests potential overlap between SMOTE's synthetic samples for minority classes and the decision boundary of majority classes. Figure \ref{fig:dynamic_all_categories_cm} suggests misclassification of majority class samples as minority classes, but the overall model improvement with SMOTE justifies this cost.

\begin{figure*}
  \centering

  \subfloat[CNN]{\includegraphics[width=0.5\textwidth]{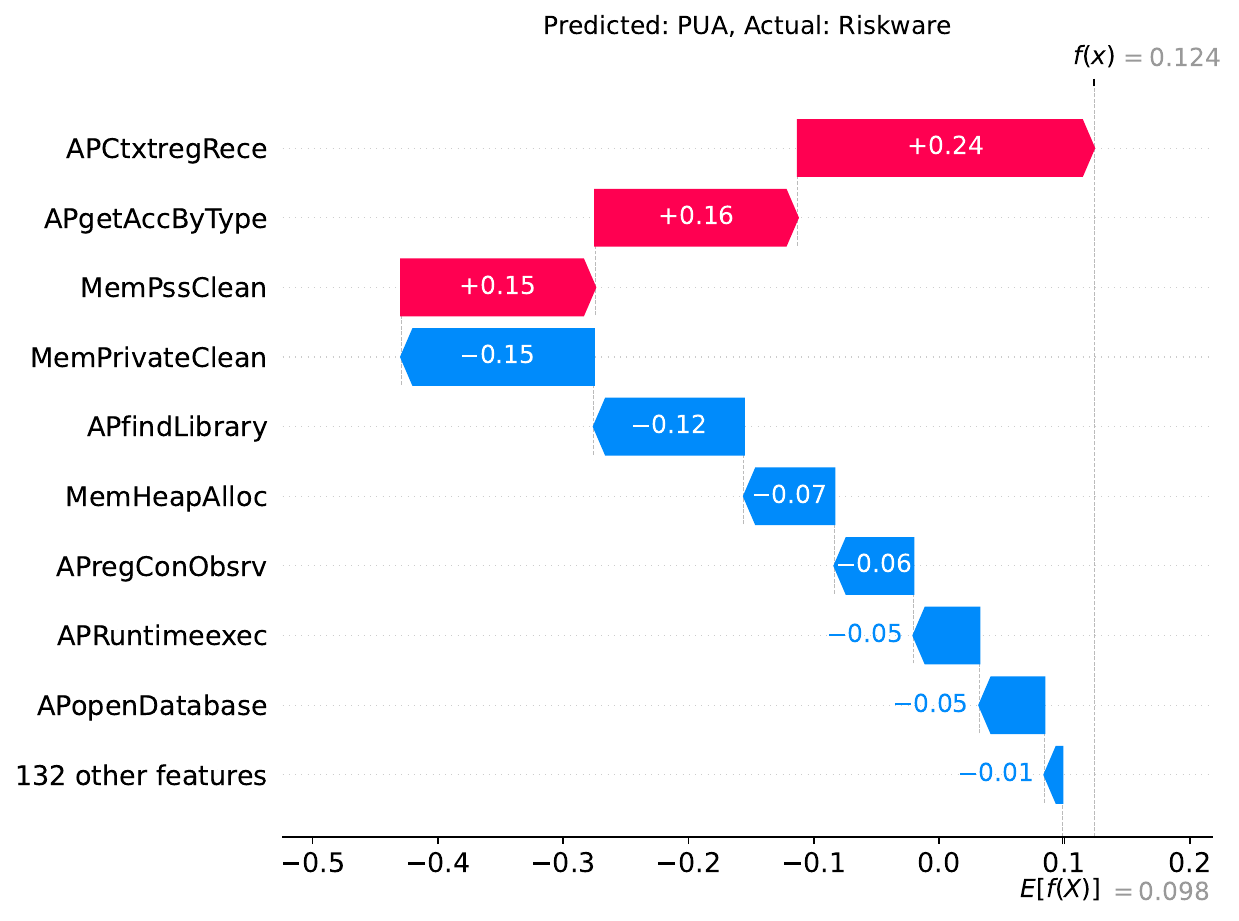}}\hfill
  \subfloat[FFNN]{\includegraphics[width=0.5\textwidth]{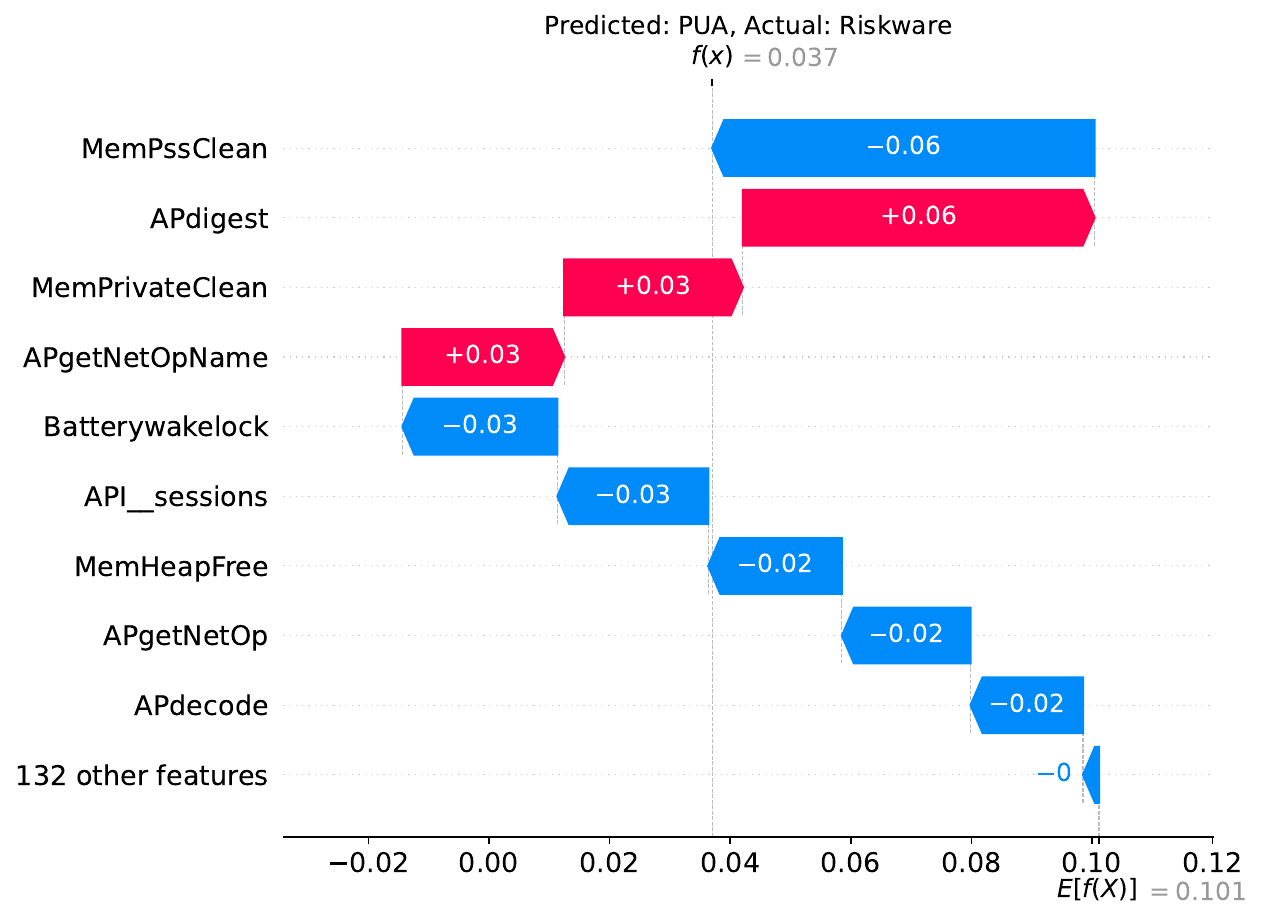}}

  \caption{Waterfall plots - Local interpretations of misclassified Riskware sample of dynamic data set}
  \label{fig:dynamic-riskware-waterfall-plots-m}
\end{figure*}

\begin{figure*}
  \centering

  \subfloat[CNN]{\includegraphics[width=0.5\textwidth]{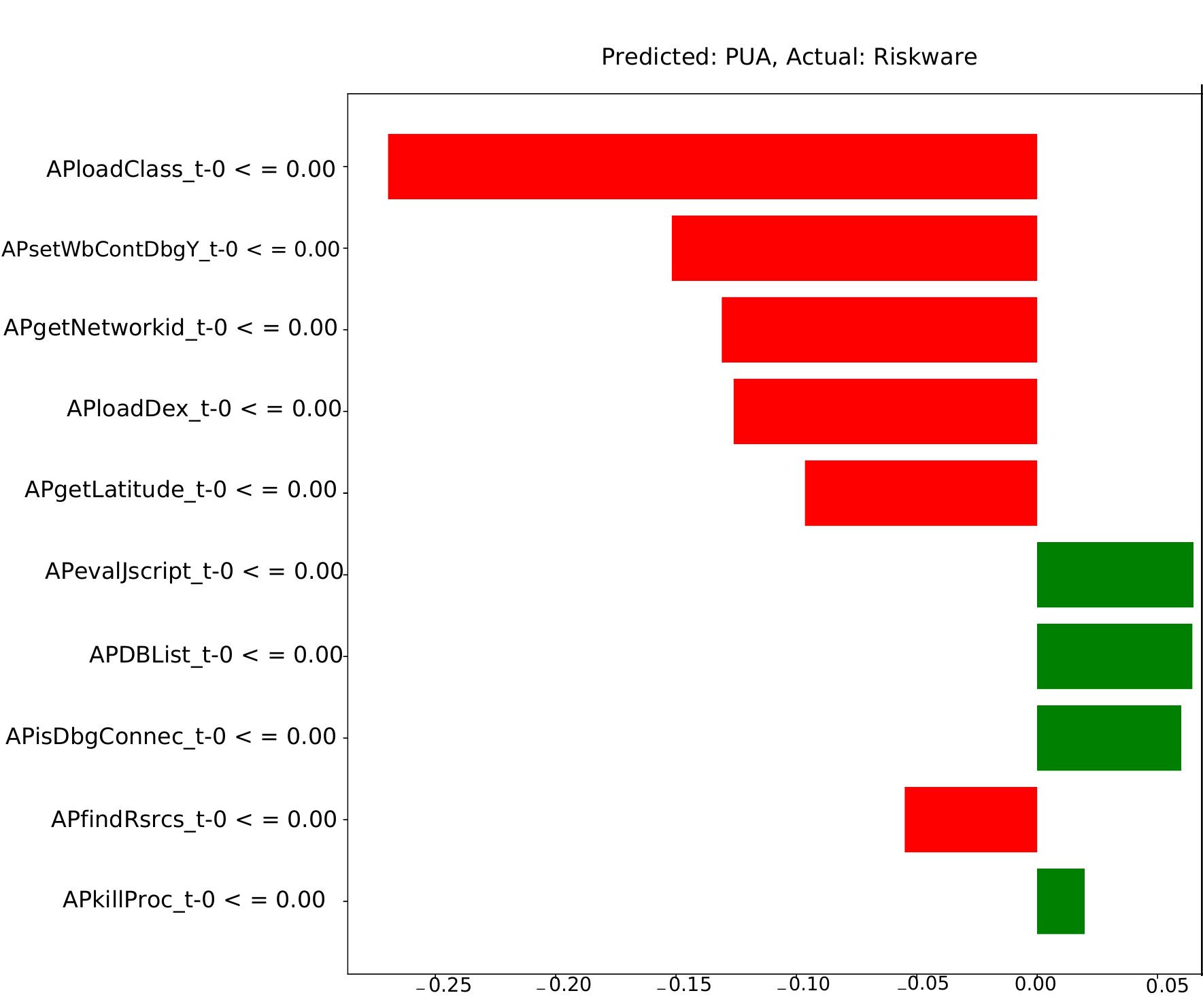}}\hfill
  \subfloat[FFNN]{\includegraphics[width=0.5\textwidth]{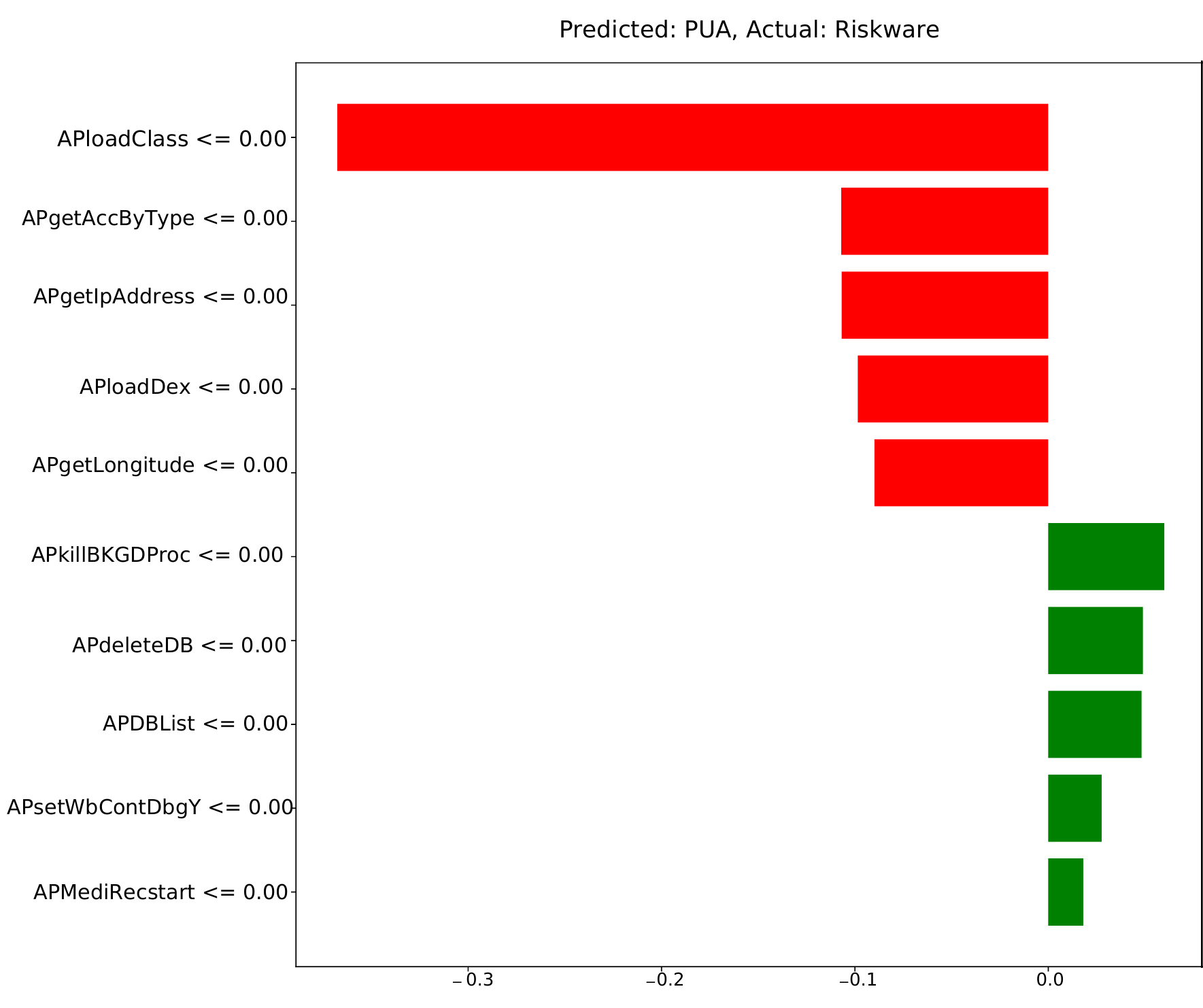}}

  \caption{LIME plots - Local interpretations of misclassified Riskware sample of dynamic data set}
  \label{fig:dynamic-riskware-lime-m}
\end{figure*}

\begin{figure*}
  \centering

  \subfloat[CNN]{\includegraphics[width=0.5\textwidth]{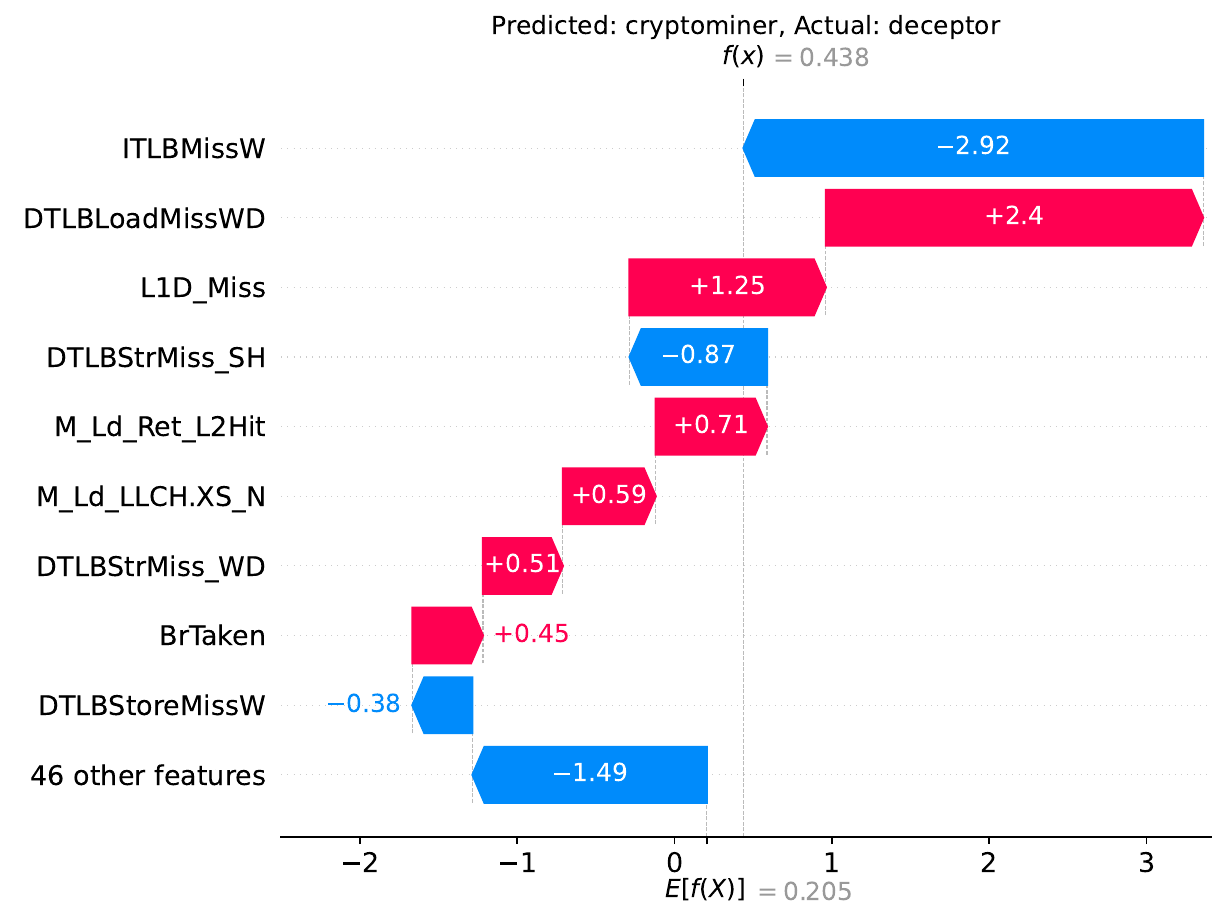}}\hfill
  \subfloat[FFNN]{\includegraphics[width=0.5\textwidth]{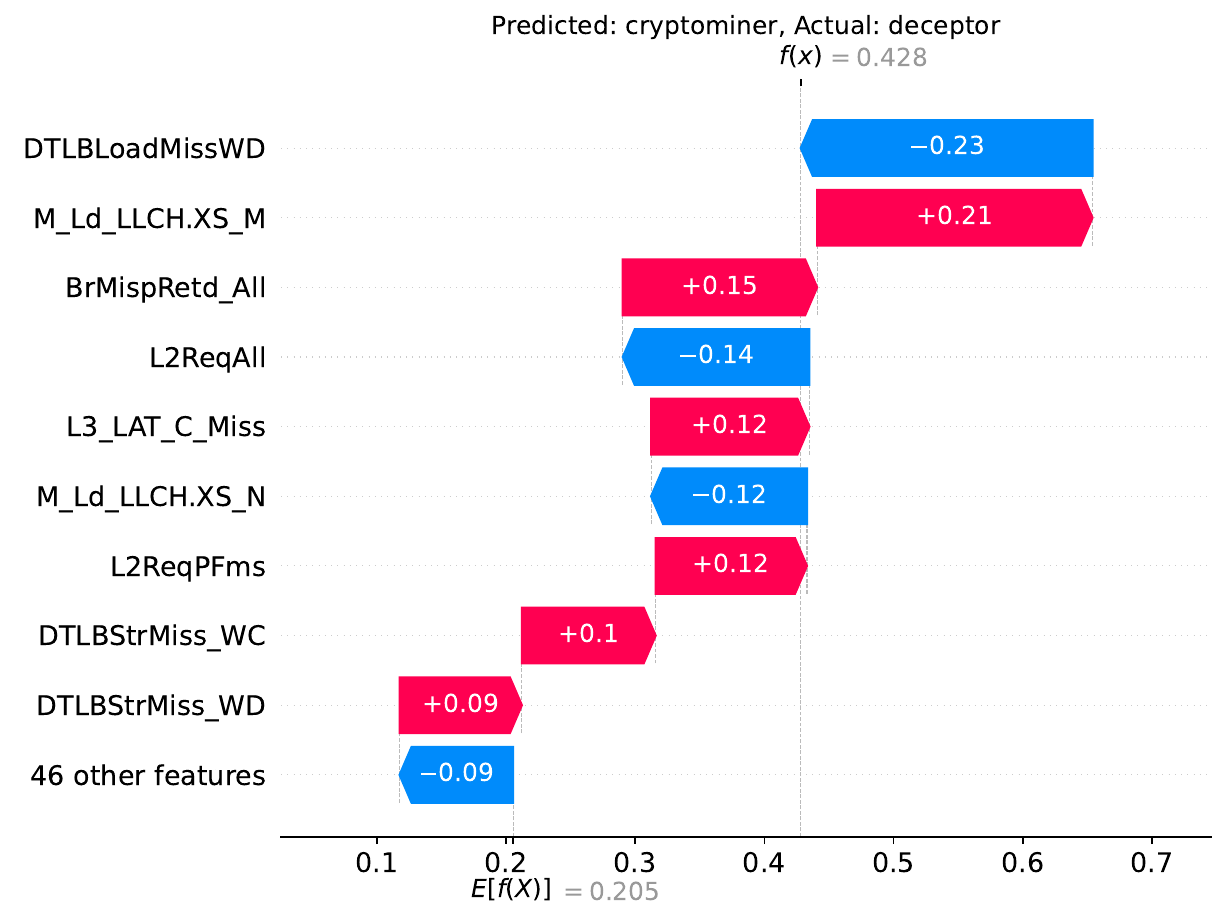}}

  \caption{Waterfall plots - Local interpretations of misclassified Deceptor sample of online data set}
  \label{fig:online-deceptor-waterfall-plots-m}
\end{figure*}

\begin{figure*}
  \centering

  \subfloat[CNN]{\includegraphics[width=0.5\textwidth]{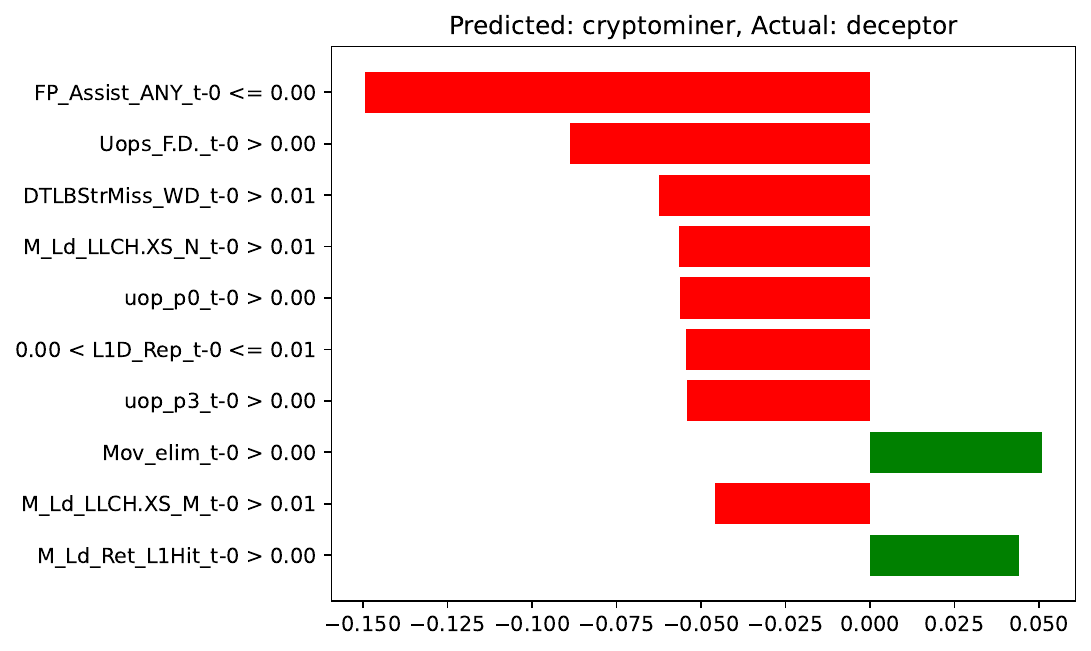}}\hfill
  \subfloat[FFNN]{\includegraphics[width=0.5\textwidth]{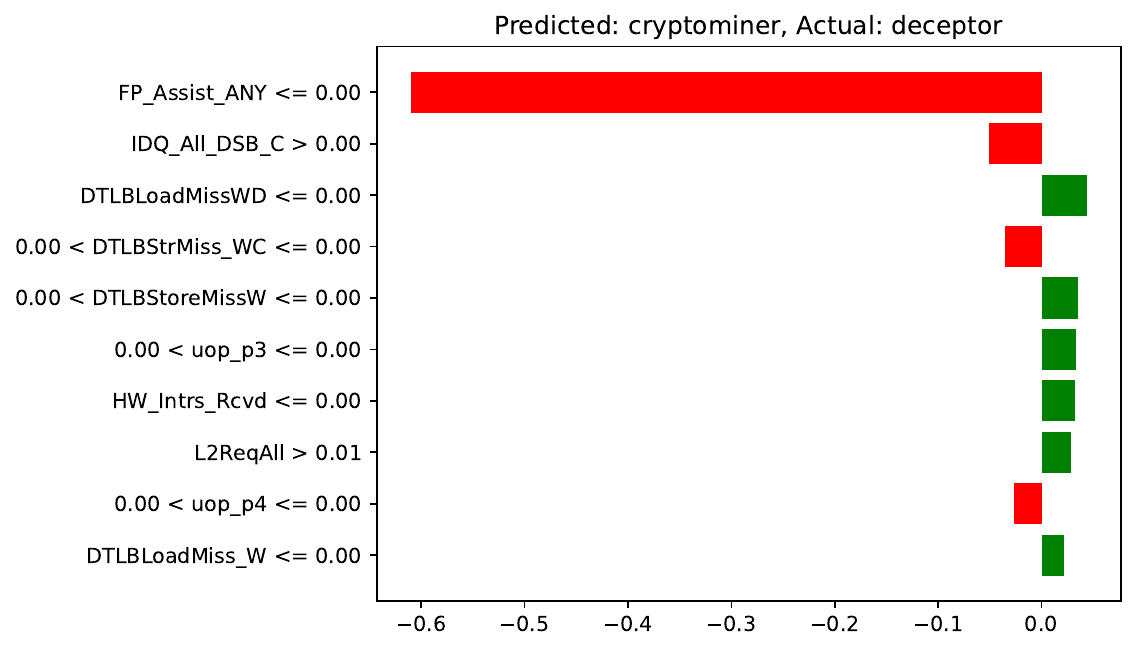}}

  \caption{LIME plots - Local interpretations of misclassified Deceptor sample of online data set}
  \label{fig:online-deceptor-lime-m}
\end{figure*}

Table \ref{tab:online-performance} presents performance metrics for the online analysis before and after SMOTE intervention. The FFNN achieved an F1 score of 78.79\%, while the CNN achieved 85.63\%. The CNN outperformed the FFNN, possibly due to its complexity, yet both models underperformed compared to those in the dynamic analysis. This could be attributed to the loss of time-series data and the introduction of synthetic samples via SMOTE, as depicted in Figure \ref{fig:online_all_categories}. We focused our explanation analysis on the Riskware class for the dynamic dataset and the Deceptor class for the online dataset due to their low F1 scores, indicating the need for further investigation. More analyses for other classes are available on our GitHub repository\footnote{GitHub: https://github.com/SecurityCard/explainability-graphs.git}.

\subsection{Global Explanation}
With reduced sample sizes, SHAP's \textit{DeepExplainer} computed SHAP values in about 6 minutes for the CNN and 16 seconds for the FFNN on the dynamic dataset. The summary plots in Figure \ref{fig:dynamic_global_shap} illustrate feature importance, with features arranged by their effects' magnitudes across all classes. Notably, API calls and Memory features were top features for both models, with {\itshape API\_\_sessions} particularly significant for the CNN model. Permutation Importance, validating SHAP's effectiveness, identified top features after permutation. The FFNN runtime was just under an hour, and for the CNN, two hours. Table \ref{tab:dynamic-permutation-importance} lists top features and their mean importance when permuted, with 10 features shared with SHAP for the FFNN and 7 for the CNN, confirming a balanced trade-off between computation time and SHAP accuracy.

For the online dataset, \textit{DeepExplainer} computed SHAP values in about 92 seconds for the CNN and 4 seconds for the FFNN. Permutation Importance took 40 minutes for the FFNN and an hour for the CNN. Figure \ref{fig:online_global_shap} and Table \ref{tab:online-permutation-importance} reveal that among the globally identified top 10 features by Permutation Importance, 5 are within the top 10 identified by SHAP for the FFNN, and 6 for the CNN, indicating a balanced trade-off between computational accuracy and resource costs. Additionally, 4 features were shared among those identified by SHAP for both models, suggesting robustness in the online analysis models. However, this might not be as pronounced as in the dynamic analysis due to limitations with SHAP's handling of time-series-based neural networks.

\subsection{Local Explanation}
For local explanations, SHAP generated waterfall plots, focusing on misclassified samples for the classes with the lowest F1 score. Figure \ref{fig:dynamic-riskware-waterfall-plots-m} displays the waterfall plots for a dynamic sample misclassified by both models, showcasing SHAP values for each feature. Positive contributions (in pink) push the probability towards the predicted class, while negative contributions (in blue) push it away. The sum of SHAP values for each sample represents the difference between the base value $E[f(x)]$ at the bottom and the final output value $f(x)$ at the top. For Figure \ref{fig:dynamic-riskware-waterfall-plots-m}, the final output values indicate probabilities for the predicted class (PUA) and not the actual class (Riskware), being 0.124 and 0.037 for the CNN and FFNN, respectively. These scores while still positive, indicating the result of being classified as PUA, are not large quantities, meaning the classifier decision is not as certain. Once again the majority of the features are either of the Memory or the API types, with the respective model explanations even sharing two features, which are also globally important to model predictions. Figure \ref{fig:dynamic-riskware-lime-m} shows LIME graphs for the same misclassified sample, providing different feature importance compared to SHAP. Green bars (right-directed) indicate positive contributions towards the predicted class, while red bars (left-directed) suggest negative contributions for all classes not predicted. The LIME explanations indicate most of the decision to misclassify was due to negative feature contribution, supporting our conclusion from the waterfall plot analysis about the classifier decision not being certain. The top 10 features identified by LIME are all of the API type. This further indicates that it may be possible to increase model performance by removing the features that are less relevant to model decision making, which has significant implications for future research and for increasing the effectiveness of real-world remediation strategies.

For online analysis, Figure \ref{fig:online-deceptor-waterfall-plots-m} displays waterfall plots for a misclassified Deceptor sample, showing similar characteristics to dynamic local explanations with classifier decision not being as certain and many features having equal and opposite contributions to each other. Figure \ref{fig:online-deceptor-lime-m} depicts LIME graphs for the same misclassified sample, with shared top two features providing strong negative impacts on classifying this sample as not the other classes. By combining the locality specific explainability methods of Permutation Importance and LIME to the both global and local method of SHAP, we've effectively reached an even deeper understanding and interpretation of these black-box models for both dynamic and online malware classification. This method of combining different explanaibility methods has significant implications for future works that even just choose to hone in on dynamic or online malware category classification.

\section{Conclusion and Future Work}\label{sec:conclusion}
This paper applies FFNN and CNN models to dynamic and online malware datasets for classification, using SHAP, LIME, and Permutation Importance for explanations. This approach balances resource costs and analysis depth. Despite class imbalance in the dynamic dataset, SMOTE partially mitigates this issue, though with potential performance degradation. Another limitation arises from explainability methods unsuitable for time-series data. Future research should explore this limitation and evaluate model performance on diverse datasets.

Future work aims to investigate adversarial attacks' potential exploitation by malicious users to misclassify malware categories. By focusing on features of highest importance identified by explainability methods, our findings can contribute to creating more efficient datasets, facilitating real-world malware remediation and aiding cyber-analysts in coordinating responses to threats.
\section*{Acknowledgment}
This work is supported by the NSF Scholarship for Service Program Award 2043324 and 2230609.

\bibliographystyle{IEEEtran}
\bibliography{references}

\end{document}